\documentclass[a4paper]{article}
\newcommand{\affil}[1]{$^{\rm #1}$}
\usepackage[authoryear]{natbib}
\bibpunct{(}{)}{;}{a}{}{,}
\usepackage{graphicx}
\date{} 
\title{\large\bf\flushleft 
Empirical evidences in favor of a varying-speed-of-light}

\author{\parbox{\textwidth}{\flushleft
\vspace{-0.5cm}
{\it Yves-Henri Sanejouand\affil{A,B}}\\
\vspace{0.4cm}
{\small \affil{A}\,Laboratoire U3B, UMR 6204 of CNRS, Facult\'e des Sciences, 2 rue de la Houssini\`ere, 44322 Nantes Cedex 3, France.
}\\
{\small \affil{B}\,Email: Yves-Henri.Sanejouand@univ-nantes.fr}}}
\begin{document}
%
\begin{changemargin}{.8cm}{.5cm}
\begin{minipage}{.9\textwidth}
\vspace{-1cm}
\maketitle
%
%
\small{\bf Abstract: Some empirical evidences in favor of the hypothesis that the speed of light
decreases by a few centimeters per second each year are examined. Lunar laser ranging data are found to be 
consistent with this hypothesis, which also provides a straightforward
explanation for the so-called Pioneer anomaly, that is, a time-dependent blue-shift observed when analyzing
radio tracking data from distant spacecrafts, as well as an alternative explanation for both the apparent
time-dilation of remote events and the apparent acceleration of the Universe.
The main argument against this hypothesis, namely, the constancy of fine-structure and Rydberg
constants, is discussed. Both of them being combinations of several physical constants, their constancy
imply that, if the speed of light is indeed time-dependent, then at least two other ``fundamental constants'' have to vary as well.
This defines strong constraints, which will have to be fulfilled by future varying-speed-of-light theories. 
}

\medskip{\bf Keywords:}
Lunar laser ranging --
Lenght of day -- 
Pioneer anomaly --
Time dilation --
Supernovae --
Hubble's law --
Cosmological constant --
Fine Structure constant --
Rydberg constant. 

\medskip
\medskip
\end{minipage}
\end{changemargin}
\small

\section{Introduction}

During the twentieth century, the speed of light has reached the theoretical status
of a ``universal constant'', a fixed value of  $c_0=$ 299,792,458 m s$^{-1}$ being chosen
in 1983 as a basis for the international unit system. 

\noindent
In the present study, empirical evidences in favor of the hypothesis
that the speed of light actually varies as a function of time are examined. 
It is by far not the first attempt to put forward such an hypothesis \citep{Wold:35, North, Magueijo:00}
but it is only recently that measurements accurate enough,
on periods of time long enough, have allowed to witness several independent phenomenons
in rather good agreement with it.

\section{Main hypothesis}

It is assumed herein that $c(t)$, the time-dependent speed of light, varies slowly on
the considered timescales, so that it can be approximated by:
\[                    
c(t) = c_0 + a_c t + \frac{1}{2} \dot{a}_c t^2 +
\cdots 
\]
where $a_c$ is the time derivative of $c(t)$, 
$\dot{a}_c$ the time derivative of $a_c$, and where $c_0$ is the value of the speed
of light at $t=t_0=0$, {\it e.g.} when a series of measurements begins.
Hereafter, for the sake of simplicity, only the two first terms of this expansion are retained.
In other words, as proposed long ago \citep{Wold:35},
it is assumed that $c(t)$ varies so slowly that it can be well
approximated by:                   
\begin{equation}
\label{cdet}   
c(t) = c_0 + a_c t  
\end{equation}

\newpage 
\section{Lunar laser ranging \label{llr}}

Thanks to reflectors left on 
Moon by Apollo and Lunokhod missions, using laser pulses,
highly accurate measurements of $\delta t_M$, the time 
taken by light to go to the Moon and back to
Earth, have been performed over the last fourty years \citep{Dickey:94}.

\noindent
If $d_M$, the average Moon semi-major axis, is assumed to have {\it not}
significantly changed over this timespan, then, as a consequence of (\ref{cdet}):
\[
\delta t_M = \frac{2 d_M}{c(t)}
\]
is expected to vary as a function of time, so that: 
\begin{equation}
\label{dtl} 
\dot{\delta t_M} =  \frac{-2 a_c d_M}{c_0^2}
\end{equation}
As a matter of fact, a value of $\dot{\delta t_M}= 0.255 \pm 0.005$ nsec per year has been measured \citep{Dickey:94}.
According to  (\ref{dtl}), this yields $a_c = -9.4\ 10^{-10}$m s$^{-2}$.

\noindent
Since, nowadays, it is assumed that $c(t) = c_0$, the increase of $\delta t_M$ is usually interpreted as 
an increase of $d_M$, of 3.82 $\pm$ 0.07 cm per year \citep{Dickey:94}.
Such a steadily increase calls for an explanation, which is usually given as follows:
since the Moon exerts a gravitational torque on the bulge of Earth, energy dissipation due to tidal friction
yields a decrease of Earth rotation rate, which corresponds to a
secular increase of the length of the day (LOD).
In turn,
as a consequence of momentum conservation, 
the Earth-Moon distance has to increase as well \citep{Darwin:1879}.
But, in order to account for an increase of $d_M$ of 3.8 cm per year, $\dot T_{LOD}$,
the increase of LOD, has to be of 2.3 msec cy$^{-1}$ \citep{Stephenson:95}, while 
current estimates are significantly smaller. 
Indeed, paleotidal values provided by late Neroproterozoic tidal rhythmites yield an average
of $\dot T_{LOD}=$ 1.3 msec cy$^{-1}$ over the last 620 million years \citep{Williams:00},
in close agreement with the value obtained by analyzing paleoclimate records 
over the last 3 million years, namely, 1.2 msec cy$^{-1}$ \citep{Lourens:01}.
Although,
using an extensive compilation of ancient eclipses,
a larger value  
of 1.70 $\pm$ 0.05 msec cy$^{-1}$ for the last 2500 years has been obtained \citep{Stephenson:95},
it may prove to be not that relevant on other timescales since, for instance,
fluctuations of several milliseconds have been observed over the last centuries,
likely to be due to mantle-core interactions \citep{Eubanks:93} or events like
the warm El Nino Southern Oscillation, which is accompanied
by an excess in atmospheric 
angular momentum \citep{Munk:02}.
\begin{figure}[t]
\begin{center}
 \vskip -1mm
 \includegraphics[width=10.0cm]{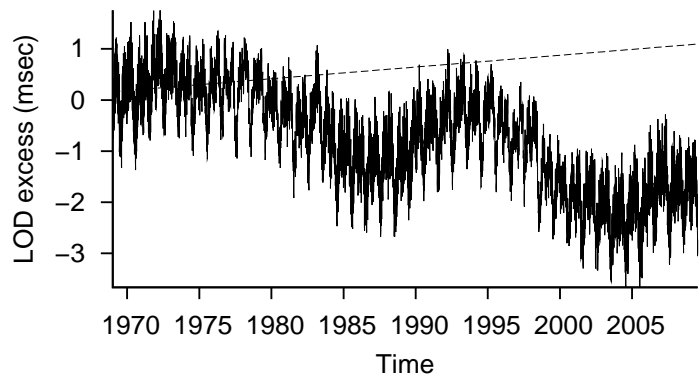}
 \vskip -1mm
 \caption{\small Length of day since 1969, that is, since Earth-Moon laser ranging data started to be collected.
The dotted line shows the 2.3 msec cy$^{-1}$ trend expected as a consequence of momentum conservation,
when it is assumed that, using laser ranging, an actual increase
of Earth-Moon distance is measured.
LOD data comes from the EOP 05 CO4 series \citep{Gambis:09}, as provided by the Earth Orientation Centre
(http://hpiers.obspm.fr).}
\label{LOD69}
\end{center}
\end{figure}
As a matter of fact, as shown in Fig. \ref{LOD69}, since 1969,
that is, since lunar laser ranging data have started to be collected,
the mean LOD has {\it decreased} \citep{Gambis:00}.
So, it seems likely that at least half of  
the observed increase of $\delta t_M$ is {\it not} due to tidal forces. Instead, it could well 
indicate an actual decrease of the speed of light.

\newpage
\section{Pioneer anomaly \label{Pioneer}}

A straightforward way to check this later hypothesis is to emit
an electromagnetic wave with a given frequency $\nu_0$, and then to
measure its wavelength as a function of time:
\[
\lambda_{mes} = \frac { c(t) }{\nu_0} 
\]
since, as a consequence of  (\ref{cdet}), it should drift according to: 
\begin{equation}
\label{wldr} 
\lambda_{mes} = \lambda_0  ( 1 + \frac { a_c}{c_0} t )
\end{equation}
which, with $a_c < 0$, means that a blue-shift increasing linearly in time
should be observed as if, when interpreted as a Doppler effect,
the source were accelerating towards the observer.

\noindent
As a matter of fact, such a time-dependent blue-shift may well have already been observed,
by analyzing radio tracking data from Pioneer 10/11 spacecrafts \citep{Anderson:98}.
During this series of experiments, a signal was emitted towards the spacecraft, 
up to 67 astronomical units (AU) away, at $\nu_0$ = 2.292 GHz,
using a digitally controlled oscillator,
sent back to Earth by the spacecraft transponder where $\lambda_{mes}$, 
the wavelength of the radiometric photons received\footnote{Since the Pioneer spacecrafts beamed their
radiometric signal at a power of eight watts \citep{Anderson:02}, photons are indeed detected one-by-one
when the anomaly is exhibited, both spacecrafts being more than 10-15 AU away from the Sun \citep{Anderson:02}},
was measured with the antenna complexes and the low-noise maser amplifiers of the 
Deep Space Network \citep{Anderson:02}.
An apparent anomalous, constant, acceleration, $a_p$, roughly directed towards the Sun
was left unexplained, with $a_p = 8.74 \pm 1.33 \ 10^{-10}$m s$^{-2}$ \citep{Anderson:98},
in spite of extensive attempts to unravel its physical nature \citep{Anderson:02, Anderson:05}.
In particular, it is unlikely to have a gravitational origin since such a constant acceleration, 
on top of Sun's attraction,
would have been detected when analyzing orbits within the Solar System, 
noteworthy for Earth and Mars \citep{Anderson:98}
but also for, {\it e.g.}, trans-neptunian objects \citep{Wallin:07}.
On the other hand, although directed heat radiation may well play 
a role \citep{Anderson:98,Anderson:02}, it seems unlikely to account for
more than 25\% of the measured effect \citep{Anderson:09}.

\noindent
Moreover, this anomaly was confirmed by at least two other independent analyses of the data,
providing similar estimates for the effect, 
namely  $a_p = 8.60 \pm 1.34 \ 10^{-10}$m s$^{-2}$ \citep{Markwardt:02}
and $a_p = 8.4 \pm 0.1 \ 10^{-10}$m s$^{-2}$ \citep{Reynaud:09}.
Interestingly, this later study confirmed that small amplitude, periodic variations, 
of the anomaly do occur \citep{Anderson:02},
the main component period being equal to Earth's sidereal rotation
period, while a semi-annual component of similar magnitude  
is also exhibited \citep{Reynaud:09},
corresponding to a fluctuation of $\approx 0.2 \ 10^{-10}$m s$^{-2}$ \citep{Anderson:05pb}.

\noindent
These results are in good agreement with our hypothesis.
Indeed, $a_c = - a_p$  yields a value for $a_c$ which 
would explain 90\% of the increase of  $\delta t_M$, while
assuming that $a_p$ is directed towards the Sun, if it actually happens to be an apparent effect and,
as such, is instead directed towards the Earth, 
introduces the following artefactual, periodic, fluctuation:
\[
a_{mes} = a_p cos \alpha
\]
where $a_{mes}$ is the value taken into account in models used for interpreting the radiometric data 
\citep{Anderson:02}
and where $\alpha$ is the angle between spacecraft to Sun and spacecraft to Earth directions. 
When the spacecraft is far away from the observer, 
$\delta a_p$, the amplitude of the corresponding fluctuation, 
is approximately given by:
\[
\delta a_p = \frac{1}{2} a_p \frac{d_E}{d_p}
\]
where $d_p$ is the Earth-spacecraft distance, $d_E$ being Earth semi-major axis. 
So, according to our hypothesis, 
the amplitude of the semi-annual periodic component of $a_p$ is expected to
ly in a range including the reported value, since
it has to decrease from $\delta a_p \approx 0.3~10^{-10}$m s$^{-2}$,
when the spacecraft is 15 AU away from the Sun, that is, when the anomaly starts
to show up in the radiometric data \citep{Anderson:02}, to $\delta a_p \approx 0.1~10^{-10}$m s$^{-2}$,
when the spacecraft is 67 AU away from the Sun, that is, when the last useful data from Pioneer 10 were
collected.  

\newpage
\section{Time dilation \label{sec:tdil}}

Both previous estimates of $a_c$ (see Table \ref{summary}) come from measurements
performed within the Solar System, over a few decades. However, it has been noticed, then as a
numerical coincidence, that 
$a_p$ is nearly equal to $H_0 c_0$, where $H_0$ is the Hubble constant \citep{Anderson:02}.
Within the frame of the present study, this suggests that the decrease of the speed of light
at a rate of the order of magnitude of $a_c$ 
may be revealed by studying phenomenons occuring over cosmological distances. 

\noindent
Indeed, as a consequence of  the decrease of the speed of light,
the timescale of remote events, for instance, is expected to be overestimated.
To exhibit this effect in a clear-cut way,
let us consider the case of a {\it static} Universe.
Then, when two signals are emitted at times $t_i$ and $t_j$,
$L$, the distance between both is: 
$L = c(t_{em}) T_0$, if it is assumed that 
during $T_0 =  t_j - t_i$ the speed of light at $t=t_{em}$, $c(t_{em})$, 
does not change significantly.
On the other hand,
since (\ref{cdet}) is {\it not} spatially dependent,
$L$ is expected to
remain constant during the flight of the signals towards the observer
who measures $T_{mes}$, the time delay between both, as:
\[
T_{mes}= \frac {L}{c_0}
\]
that is:
\[
T_{mes}= \frac {c(t_{em})}{c_0} T_0
\]
With  (\ref{cdet}), this yields:
\begin{equation}
\label{tdil} 
\frac {T_{mes}}{T_0}= 1 - \frac { a_c \Delta t_g}{c_0}
\end{equation}
where $\Delta t_g = t_{em} - t_0$ is the photon time-of-flight between the source and the observer
and where the minus sign comes from the fact that $c_0$ is the value of the speed of light
when the observation is performed. 

\noindent
On the other hand, up to ten years ago, $z$, the redshift of a galaxy, had been shown to be well described as a linear
function of $d_g$, its distance, such that:
\begin{equation}
\label{hubble} 
z = \frac {H_0 d_g}{c_0}  
\end{equation}
However, as an empirical law,  (\ref{hubble}) can also
be written in the following form:
\begin{equation}
\label{zdt} 
z = H_0 \Delta t_g
\end{equation}
while with  (\ref{zdt}),  (\ref{tdil}) becomes:
\begin{equation}
\label{tdez} 
\frac {T_{mes}}{T_0}= 1 - \frac{ a_c}{H_0 c_0} z
\end{equation}

\noindent
As a matter of fact, it has been observed that
light curves of distant supernovae are dilated in time
\citep{Hamuy:96b,Riess:96},
according to: 
\begin{equation}
\label{aging}
\frac {T_{mes}}{T_0} = 1 + z 
\end{equation}
where $T_0$ and $T_{mes}$ are the typical timescales of the event, as observed in
the case of nearby and distant supernovae, respectively.
Indeed, nowadays, a stretching by a $(1 + z)$ 
factor of reference, nearby supernovae, light curves is included
in all analyses of distant supernovae \citep{Riess:04, Riess:07}.

\noindent
Such a phenomenon can be understood
within the frame of standard cosmological models \citep{Schrodinger:39,Wilson:39}. 
However, if it is assumed that the decrease of the speed of light is responsible for
most of this effect, (\ref{tdez}) and (\ref{aging}) yield:
\begin{equation}
\label{hc}
a_c = -H_0 c_0
\end{equation}
Note that with $H_0 = 72 \pm 8$ km s$^{-1}$ Mpc$^{-1}$ \citep{Freedman:01},
$a_c=-7.0 \pm 0.8\ 10^{-10}$m s$^{-2}$, that is, a value in the range of both previous estimates
(see Table \ref{summary}).

\begin{figure}[t]
\begin{center}
 \vskip +5mm
 \hskip -5mm
 \includegraphics[width=8 cm]{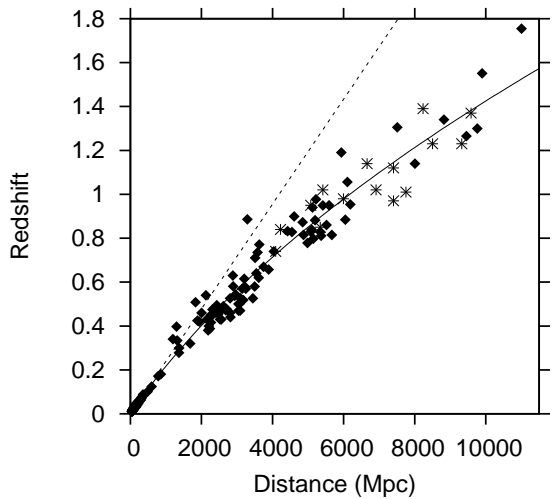}
 \vskip +1mm
 \caption{\small Redshift of type Ia supernovae as a function of their distance, in megaparsecs.
 Black diamonds: the 156 supernovae of the initially compiled "gold set" \citep{Riess:04}.
Stars: 16 cases observed more recently, with the Hubble Space Telescope \citep{Riess:07}.
 Dotted line: Hubble's law. Plain line: a single-parameter fit of the data,
performed using the relationship obtained following
the varying-speed-of-light hypothesis
 discussed in the present study.}
\label{sn07}
\end{center}
\end{figure}

\newpage
\section{Supernovae redshifts}

Moreover, under the additional, rather natural, hypothesis 
that (\ref{zdt}) is a more generally valid form of Hubble's law than (\ref{hubble}), 
as a consequence of 
the time-dependence of the speed of light, $z$ is expected to be a non-linear function of  $d_g$.
Indeed, using (\ref{cdet}), one gets: 
\[
d_g = c_0 \Delta t_g - \frac {1}{2} a_c \Delta t_g^2 
\]
This yields:
\begin{equation}
\label{dt}
\Delta t_g = \frac {c_0} {a_c}
(1 - \sqrt{ 1 - \frac {2 a_c d_g}{c_0^2}})
\end{equation}
and, with (\ref{zdt}):
\begin{equation}
\label{zdnew}
z = \frac {H_0 c_0}{a_c} (1 - \sqrt{ 1 - \frac {2 a_c d_g}{c_0^2}})
\end{equation}
which, for short distances, 
can be approximated by (\ref{hubble}).

\noindent
As a matter of fact, using the rather homogeneous type Ia
supernovae subclass (Sne Ia) as standard candles, it was shown that,
for large values of $d_g$,  
Hubble's law is not linear any more \citep{Riess:98,Perlmutter:99}.
This is illustrated in Fig. \ref{sn07} for a ``gold set'' of 182 Sne Ia \citep{Riess:04,Riess:07},
together with a least-square fit performed with  (\ref{zdnew}) 
which yields $a_c = -6.6\ 10^{-10}$m s$^{-2}$.
In practice, distances are obtained from extinction-corrected distance moduli,
$m - M = 5\ log_{10}d_g + 25$, where $m$ and $M$ are the apparent and the absolute
magnitudes of the supernovae, respectively \citep{Riess:07}.

\noindent
The explanation nowadays given for the nonlinearity of Hubble's law rely on
an acceleration of universe's expansion due to a non-zero, although very small,
value of $\Lambda$, the cosmological constant \citep{Riess:98,Perlmutter:99}.
However,  this explanation looks like all previous
attempts to introduce a non-zero $\Lambda$ in the equations of General Relativity,
namely, like an {\it ad hoc} one. Indeed, $\Lambda$ was first added into these equations
by Einstein himself, so as to obtain a static solution for the Universe as a whole \citep{Einstein:17},
next kept by Lemaitre, in order to account for a then too large measured value of Hubble constant,
with respect to the age of Earth \citep{Lemaitre:27}, and it has now
been reintroduced so as to explain why the Universe seems to accelerate, 
instead of the deceleration expected within the frame of standard cosmological models,
as a consequence of gravitational forces.
In all these cases, a non-zero value of $\Lambda$ allowed to rescue a theory unable to explain a 
seemingly obvious fact. 
Although $\Lambda$ helps improving the standard cosmological model, noteworthy
within the frame of the ``concordance model'', note that this is at the cost of introducing both 
a ``cosmic coincidence'' \citep{Zlatev:99}  and
a new kind of so-called ``dark energy'', of unknown origin but accounting for as much as 70\% of 
universe's energy \citep{Glanz:98,Copeland:06}. 

\noindent
Interestingly, the value of $a_c$ obtained through the present analysis
is found to be nearly equal to the previous one (see Section \ref{sec:tdil} or Table \ref{summary}).   
Indeed, assuming that (\ref{hc}) is exact,
(\ref{zdnew}) takes the
following, appealingly simple, parameter-free form:
\[
z = \sqrt{ 1 + \frac {2 H_0 d_g}{c_0}} - 1
\]
already advocated in a previous study \citep{Sanejouand:05}.
Note that since, in the case of a wave, $T_{mes}=\frac{\lambda_{mes}}{c_0}$ and
$T_0=\frac{\lambda_0}{c_0}$, while $z=\frac{\lambda_{mes}-\lambda_0}{\lambda_0}$,
(\ref{zdt}) can be obtained from (\ref{tdil}) and (\ref{hc}).
In other words, if (\ref{hc}) happens to be exact, then Hubble's law is itself
a consequence of the decrease of the speed of light, as proposed seventy four years ago \citep{Wold:35}.

\begin{table*}[t]
\begin{center}
\begin{tabular}{lcc}
  \hline
                           &                                    & \\
   Empirical fact & Implied value for $a_c$ & Comments \\
                          &       (m s$^{-2}$)         & \\
  \hline
                                      &                                    & \\
   Apparent increase of &  $-9.4 \pm 0.2 \ 10^{-10}$     &  Tidal forces are expected to be \\  
  Earth-Moon distance &                                             & partly responsible for this effect. \\
                                      &                                             & \\
   Apparent acceleration & $-8.7 \pm 1.3 \ 10^{-10}$  & \citep{Anderson:98} \\
   of Pioneer 10/11              &  $-8.6 \pm 1.3 \ 10^{-10}$  & \citep{Markwardt:02} \\
                                     &    $-8.4 \pm 0.1 \ 10^{-10}$   & \citep{Reynaud:09} \\
                                       &                                             & \\
   Apparent time dilation & $-7.0 \pm 0.8 \ 10^{-10}$  & Depends upon the actual value of $H_0$. \\ 
   of remote events          &                                           & \\
                                        &                                             & \\
   Apparent acceleration & $-6.6 \pm 0.7 \ 10^{-10}$  & Depends upon the actual value of $H_0$. \\ 
   of the Universe          &                                           & \\
                                      &                                    & \\
   \hline
 \end{tabular}
 \caption{\small Values obtained for $a_c$, the rate of change of the speed of light,
through the analysis of four different kinds of experimental data, 
collected over two widely different timescales, namely, decades (top) and billions of years (bottom).
$H_0$ is the Hubble constant. 
  } \label{summary}
 \end{center}
 \end{table*}

\section{Fine-structure constant}

The speed of light plays a pivotal role in many physical phenomenons and, as such,
its variations, even at a slow rate, are expected to have far reaching consequences. Noteworthy,
$c_0$ is involved in several key combinations of physical constants,
some of which are known with high accuracy. 
In particular, this is the case of $\alpha$, the fine-structure constant:
\[
\alpha = \frac{e^2}{4 \pi \epsilon_0 \hbar c_0}
\]
where $e$ is the electron charge, $\hbar$, the Planck constant and $\epsilon_0$,
the vacuum permitivity. Indeed, it has been shown that $\alpha$
depends little upon the redshift \citep{Webb:99}, 
if it does at all \citep{Uzan:03, Petitjean:04}. 

\noindent
However,  $\alpha$ may prove constant in spite of the time dependence of
the speed of light if at least one among the other ``fundamental constants'' involved in $\alpha$
exhibits a complementary time-dependence. In the case of $\alpha$,
an obvious candidate is the vacuum permitivity since it is already known to be 
related to the speed of light, namely through:
\[
c_0 = \frac{1}{\sqrt{\mu_0 \epsilon_0}}
\]
where $\mu_0$ is the vacuum permeability. Thus, the constancy of $\alpha$ would
mean that $Z_0$, the characteristic impedance of vacuum:
\[
Z_0 = \frac{1}{ \epsilon_v(t) c(t) }
\]   
is the relevant fundamental constant, $\epsilon_v(t)$ being the time dependent vacuum permitivity. 

\noindent  
Note that it has recently been proposed to redefine the international unit system
so as to fix the values of both $K_J$ and $R_K$, the Josephson and von Klitzing constants.
Within such a frame, quantities that are nowadays assumed to have, {\it par d\'efinition},
fixed values, like vacuum permitivity and permeability, 
would become again quantities that have to be determined by experiment \citep{Mills:06}.
Note also that if both $\alpha$ and $Z_0$ are actual fundamental constants, then 
it has to be the case for
$R_K$, since $R_K = \frac{Z_0}{2 \alpha}$. 
Interestingly, like $\alpha$, $R_K$ belongs to the small set of
physical constants known with a very high accuracy \citep{Mohr:08}. 

\section{Rydberg constant}

However, introducing a time dependent vacuum permitivity \citep{Sumner:94}
should not prove enough, since there is another combination of physical constants
nowadays known to be {\it not} time-dependent, namely, 
$R_y$, the Rydberg constant \citep{Peik:06}:
\[
R_y = \frac{ m_0 c_0^2}{4 \pi \hbar } \alpha^2
\]
where $m_0$ is the electron mass. 
Likewise, the hypothesis of a varying speed of light would also prove consistent with this empirical fact
if at least another ``fundamental constant'' is actually
time-dependent, namely, either $m_0$ or $\hbar$. Since $\hbar$ is also involved in $\alpha$,
the additional hypothesis that the electron mass is time-dependent would be the simplest one.

\section{Discussion and Conclusion}

In the present study, four different kinds of experimental data have been shown to be consistent with
the hypothesis that the speed of light decreases as a function of time. As summarized in Table \ref{summary},
analyses of these data reveal that $a_c$, the rate of change of the speed of light, lies in a rather narrow
range, namely, between $-6.6$ and $-9.4 \ 10^{-10}$m s$^{-2}$, corresponding to a decrease
of the speed of light of 2.1-3.0 cm s$^{-1}$ per year. Note that the upper bound is likely to be overestimated
since tidal forces are also expected to bring a significant contribution to the observed increase of the time
taken by light to go to the Moon and back to Earth (Section \ref{llr}). 
Note also that the data considered herein have been collected on two widely different timescales.
Lunar laser ranging as well as Pioneer data have been determined over the last few decades, since 1969 and 1972,
respectively, while the apparent time dilation of remote events and the Universe's acceleration were exhibited
by analyzing light emitted billions of years ago, namely, by galaxies with $z \gg 0.1$ (see Fig. \ref{sn07}).  
This also suggests that $a_c$ has not changed significantly over this timespan.

\noindent
From an experimental point of view,
the main argument against the hypothesis advocated in the present study
is backed by the facts that both fine-structure and Rydberg constants have been shown
to vary little in time, if at all \citep{Peik:06}. Indeed, this puts severe
constraints on the development of a self-consistent theory in which the speed of light 
is varying in time since, as a consequence, some other ``fundamental constants'' have to vary accordingly.

\noindent
However, building such a theory is beyond the scope
of this paper. Although it could reveal
hidden links between a wide range of physical phenomenons and, as a consequence,
represents a challenge which is likely to arouse the interest of theoreticians,
such a work may well await confirmation
at the experimental level, as well as further clues, in order to be developped on firm grounds. 

\section*{Acknowledgments}
I thank William Sumner, Simon Mathieu, for stimulating discussions, 
H\'el\`ene Courtois,  Jean-Fran\c{c}ois Mangin, Jacques Poitevineau,
Louis Riofrio, Antonio Alfonso Faus and Francesco Piazza, as well as some referees,
for useful comments.


\begin{thebibliography}{}

\bibitem[Abarca~del Rio et~al., 2000]{Gambis:00}
Abarca~del Rio, R., Gambis, D., and Salstein, D.~A. (2000).
\newblock {Interannual signals in length of day and atmospheric angular
  momentum}.
\newblock {\em Annales Geophysicae}, 18(3):347--364.

\bibitem[Anderson et~al., 1998]{Anderson:98}
Anderson, J.~D., Laing, P.~A., Lau, E.~L., Liu, A.~S., Nieto, M.~M., and
  Turyshev, S.~G. (1998).
\newblock {Indication, from Pioneer 10/11, Galileo, and Ulysses data, of an
  apparent anomalous, weak, long-range acceleration}.
\newblock {\em Phys. Rev. let.}, 81(14):2858--2861.

\bibitem[Anderson et~al., 2002]{Anderson:02}
Anderson, J.~D., Laing, P.~A., Lau, E.~L., Liu, A.~S., Nieto, M.~M., and
  Turyshev, S.~G. (2002).
\newblock {Study of the anomalous acceleration of Pioneer 10 and 11}.
\newblock {\em Phys. Rev. D.}, 65(8):082004.

\bibitem[Anderson and Nieto, 2009]{Anderson:09}
Anderson, J.~D. and Nieto, M.~M. (2009).
\newblock {Astrometric Solar-System Anomalies}.
\newblock {\em arXiv gr-qc/0907.2469}.

\bibitem[Barrow and Magueijo, 2000]{Magueijo:00}
Barrow, J.~D. and Magueijo, J. (2000).
\newblock Can a changing $\alpha$ explain the supernovae results ?
\newblock {\em Ap. J.}, 532:L87--L90.

\bibitem[Bizouard and Gambis, 2009]{Gambis:09}
Bizouard, C. and Gambis, D. (2009).
\newblock {The Combined Solution C04 for Earth Orientation Parameters, recent
  improvments}.
\newblock {\em International Association of Geodesy Symposia Series}, 134:265.

\bibitem[Chand et~al., 2004]{Petitjean:04}
Chand, H., Srianand, R., Petitjean, P., and Aracil, B. (2004).
\newblock {Probing the cosmological variation of the fine-structure constant:
  Results based on VLT-UVES sample}.
\newblock {\em A\&A}, 417(3):853--871.

\bibitem[Copeland et~al., 2006]{Copeland:06}
Copeland, E., Sami, M., and Tsujikawa, S. (2006).
\newblock {Dynamics of Dark Energy}.
\newblock {\em International Journal of Modern Physics D}, 15(11):1753--1935.

\bibitem[Darwin, 1879]{Darwin:1879}
Darwin, G.~H. (1879).
\newblock {On the Precession of a Viscous Spheroid, and on the Remote History
  of the Earth}.
\newblock {\em Phil. Trans. R. Soc. London}, 170:447--530.

\bibitem[Dickey et~al., 1994]{Dickey:94}
Dickey, J.~O., Bender, P.~L., Faller, J.~E., Newhall, X.~X., Ricklefs, R.~L.,
  Riesi, J.~G., Shelus, P.~J., Veillet, C., Whipple, A.~L., Wiant, J.~R.,
  Williams, J.~G., and Yoder, C.~F. (1994).
\newblock {Lunar laser ranging - A continuous legacy of the apollo program}.
\newblock {\em Science}, 265(5171):482--490.

\bibitem[Einstein, 1917]{Einstein:17}
Einstein, A. (1917).
\newblock {Kosmologische Betrachtungen zur allgemeinen Relativitatstheorie}.
\newblock {\em Sitzungsberichte der Preussischen Akademie der Wissenschaften},
  1:l42--l52.

\bibitem[Eubanks, 1993]{Eubanks:93}
Eubanks, T. (1993).
\newblock {Variations in the orientation of the Earth}.
\newblock {\em Contributions of space geodesy to geodynamics: Earth dynamics,
  Geodyn. Ser}, 24:1--54.

\bibitem[Freedman et~al., 2001]{Freedman:01}
Freedman, W.~L., Madore, B.~F., Gibson, B.~K., Ferrarese, L., Kelson, D.~D.,
  Sakai, S., Mould, J.~R., Kennicutt, R.~C., Ford, H.~C., Graham, J.~A.,
  Huchra, J.~P., Hughes, S. M.~G., Illingworth, G.~D., Macri, L.~M., and
  Stetson, P.~B. (2001).
\newblock {Final results from the Hubble Space Telescope key project to measure
  the Hubble constant}.
\newblock {\em Ap. J.}, 553(1):47--72.

\bibitem[Glanz, 1998]{Glanz:98}
Glanz, J. (1998).
\newblock {Astronomy: Cosmic motion revealed}.
\newblock {\em Science}, 282(5397):2156--2157.

\bibitem[Hamuy et~al., 1996]{Hamuy:96b}
Hamuy, M., Phillips, M.~M., Suntzeff, N.~B., Schommer, R.~A., Maza, J., Smith,
  R.~C., Lira, P., and Aviles, R. (1996).
\newblock {The Morphology of Type IA Supernovae Light Curves}.
\newblock {\em Astron. J.}, 112(6):2438--2447.

\bibitem[Leibundgut et~al., 1996]{Riess:96}
Leibundgut, B., Schommer, R., Phillips, M., Riess, A., Schmidt, B., Spyromilio,
  J., Walsh, J., Suntzeff, N., Hamuy, M., Maza, J., Kirshner, R.~P., Challis,
  P., Garnavich, P., Smith, R.~C., Dressler, A., and Ciardullo, R. (1996).
\newblock {Time dilation in the light curve of the distant type Ia supernova SN
  1995K}.
\newblock {\em Ap. J.}, 466(1):L21--L24.

\bibitem[Lemaitre, 1927]{Lemaitre:27}
Lemaitre, G. (1927).
\newblock {Un Univers homog\`ene de masse constante et de rayon croissant
  rendant compte de la vitesse radiale des n\'ebuleuses extra-galactiques}.
\newblock {\em Annales de la Soci\'et\'e Scientifique de Bruxelles}, 47:49--59.

\bibitem[Levy et~al., 2009]{Reynaud:09}
Levy, A., Christophe, B., B{\'e}rio, P., M{\'e}tris, G., Courty, J.~M., and
  Reynaud, S. (2009).
\newblock {Pioneer 10 Doppler data analysis: Disentangling periodic and secular
  anomalies}.
\newblock {\em Advances in Space Research}, 43(10):1538--1544.

\bibitem[Lourens et~al., 2001]{Lourens:01}
Lourens, L.~J., Wehausen, R., and Brumsack, H.~J. (2001).
\newblock {Geological constraints on tidal dissipation and dynamical
  ellipticity of the Earth over the past three million years}.
\newblock {\em Nature}, 409(6823):1029--1033.

\bibitem[Markwardt, 2002]{Markwardt:02}
Markwardt, C.~B. (2002).
\newblock {Independent confirmation of the Pioneer 10 anomalous acceleration}.
\newblock {\em arXiv gr-qc/0208046}.

\bibitem[Mills et~al., 2006]{Mills:06}
Mills, I.~M., Mohr, P.~J., Quinn, T.~J., Taylor, B.~N., and Williams, E.~R.
  (2006).
\newblock {Redefinition of the kilogram, ampere, kelvin and mole: a proposed
  approach to implementing CIPM recommendation 1 (CI-2005)}.
\newblock {\em Metrologia}, 43(3):227--246.

\bibitem[Mohr et~al., 2008]{Mohr:08}
Mohr, P.~J., Taylor, B.~N., and Newell, D.~B. (2008).
\newblock {CODATA recommended values of the fundamental physical constants:
  2006}.
\newblock {\em Reviews of Modern Physics}, 80(2):633--730.

\bibitem[Munk, 2002]{Munk:02}
Munk, W. (2002).
\newblock {Twentieth century sea level: An enigma}.
\newblock {\em Proc. Natl. Acad. Sci. USA}, 99(10):6550--6555.

\bibitem[Nieto and Anderson, 2005]{Anderson:05}
Nieto, M.~M. and Anderson, J.~D. (2005).
\newblock {Using early data to illuminate the Pioneer anomaly}.
\newblock {\em Classical and Quantum Gravity}, 22(24):5343--5354.

\bibitem[North, 1965]{North}
North, J.~D. (1965).
\newblock {\em {The measure of the universe. A History of modern cosmology}}.
\newblock Oxford University Press.

\bibitem[Peik et~al., 2006]{Peik:06}
Peik, E., Lipphardt, B., Schnatz, H., Tamm, C., Weyers, S., and Wynands, R.
  (2006).
\newblock {Laboratory Limits on Temporal Variations of Fundamental Constants:
  An Update}.
\newblock {\em arXiv physics/0611088}.

\bibitem[Perlmutter et~al., 1999]{Perlmutter:99}
Perlmutter, S., Aldering, G., Goldhaber, G., Knop, R.~A., Nugent, P., Castro,
  P.~G., Deustua, S., Fabbro, S., Goobar, A., Groom, D.~E., Quimby, R., Lidman,
  C., Ellis, R.~S., Irwin, M., McMahon, R.~G., Ruiz-Lapuente, P., Walton, N.,
  Schaefer, B., Boyle, B.~J., Filippenko, A.~V., and Couch, W. (1999).
\newblock {$\Omega$ and $\Lambda$ from 42 high-redshift supernovae}.
\newblock {\em Ap. J.}, 517(2):565--586.

\bibitem[Riess et~al., 1998]{Riess:98}
Riess, A.~G., Filippenko, A.~V., Challis, P., Clocchiatti, A., Diercks, A.,
  Garnavich, P.~M., Gilliland, R.~L., Hogan, C.~J., Jha, S., Kirshner, R.~P.,
  Leibundgut, B., Phillips, M.~M., Reiss, D., Schmidt, B.~P., Schommer, R.~A.,
  Smith, R.~C., Spyromilio, J., Stubbs, C., Suntzeff, N.~B., and Tonry, J.
  (1998).
\newblock Observational evidence from supernovae for an accelerating universe
  and a cosmological constant.
\newblock {\em Astron. J.}, 116(3):1009--1038.

\bibitem[Riess et~al., 2007]{Riess:07}
Riess, A.~G., Strolger, L.~G., Casertano, S., Ferguson, H.~C., Mobasher, B.,
  Gold, B., Challis, P.~J., Filippenko, A.~V., Jha, S., Li, W., Tonry, J.,
  Foley, R., Kirshner, R.~P., Dickinson, M., MacDonald, E., Eisenstein, D.,
  Livio, M., Younger, J., Xu, C., Dahl\'en, T., and Stern, D. (2007).
\newblock {New Hubble Space Telescope Discoveries of Type Ia Supernovae at z
  $\geq$ 1: Narrowing Constraints on the Early Behavior of Dark Energy}.
\newblock {\em Ap. J.}, 659(1):98--121.

\bibitem[Riess et~al., 2004]{Riess:04}
Riess, A.~G., Strolger, L.~G., Tonry, J., Casertano, S., Ferguson, H.~C.,
  Mobasher, B., Challis, P., Filippenko, A.~V., Jha, S., Li, W.~D., Chornock,
  R., Kirshner, R.~P., Leibundgut, B., Dickinson, M., Livio, M., Giavalisco,
  M., Steidel, C., Benitez, T., and Tsvetanov, Z. (2004).
\newblock {Type Ia supernova discoveries at z $\geq$ 1 from the Hubble Space
  Telescope: Evidence for past deceleration and constraints on dark energy
  evolution}.
\newblock {\em Ap. J.}, 607(2):655--687.

\bibitem[Sanejouand, 2005]{Sanejouand:05}
Sanejouand, Y.-H. (2005).
\newblock {A simple varying-speed-of-light hypothesis is enough for explaining
  high-redshift supernovae data}.
\newblock {\em arXiv astro-ph/0509582}.

\bibitem[Schr\"odinger, 1939]{Schrodinger:39}
Schr\"odinger, E. (1939).
\newblock The proper vibrations of the expanding universe.
\newblock {\em Physica}, 6(9):899--912.

\bibitem[Stephenson and Morrison, 1995]{Stephenson:95}
Stephenson, F.~R. and Morrison, L.~V. (1995).
\newblock {Long-term fluctuations in the Earth's rotation: 700 BC to AD 1990}.
\newblock {\em Philos. Trans. R. Soc. London A}, 351:165--202.

\bibitem[Sumner, 1994]{Sumner:94}
Sumner, W.~Q. (1994).
\newblock {On the variation of vacuum permittivity in Friedmann universes}.
\newblock {\em Ap. J.}, 429(2):491--498.

\bibitem[Turyshev et~al., 2005]{Anderson:05pb}
Turyshev, S.~G., Nieto, M.~M., and Anderson, J.~D. (2005).
\newblock {Study of the Pioneer anomaly: A problem set}.
\newblock {\em American Journal of Physics}, 73:1033--1044.

\bibitem[Uzan, 2003]{Uzan:03}
Uzan, J.-P. (2003).
\newblock The fundamental constants and their variation: observational and
  theoretical status.
\newblock {\em Rev. Mod. Phys.}, 75:403--455.

\bibitem[Wallin et~al., 2007]{Wallin:07}
Wallin, J.~F., Dixon, D.~S., and Page, G.~L. (2007).
\newblock {Testing gravity in the outer solar system: Results from
  trans-Neptunian objects}.
\newblock {\em Ap. J.}, 666:1296--1302.

\bibitem[Webb et~al., 1999]{Webb:99}
Webb, J.~K., Flambaum, V.~V., Churchill, C.~W., Drinkwater, M.~J., and Barrow,
  J.~D. (1999).
\newblock Search for time variation of the fine structure constant.
\newblock {\em Phys. Rev. lett.}, 82(5):884--887.

\bibitem[Williams, 2000]{Williams:00}
Williams, G.~E. (2000).
\newblock {Geological constraints on the Precambrian history of earth's
  rotation and the moon's orbit}.
\newblock {\em Rev. Geophys.}, 38(1):37--59.

\bibitem[Wilson, 1939]{Wilson:39}
Wilson, O.~C. (1939).
\newblock {Possible Applications of Supernovae to the Study of the Nebular Red
  Shifts.}
\newblock {\em Ap. J.}, 90:634.

\bibitem[Wold, 1935]{Wold:35}
Wold, P.~I. (1935).
\newblock {On the Redward Shift of Spectral Lines of Nebulae}.
\newblock {\em Phys. Rev. E}, 47(3):217--219.

\bibitem[Zlatev et~al., 1999]{Zlatev:99}
Zlatev, I., Wang, L., and Steinhardt, P.~J. (1999).
\newblock {Quintessence, Cosmic Coincidence, and the Cosmological Constant}.
\newblock {\em Phys. Rev. let.}, 82(14):2858--2861.

\end{thebibliography}

\label{lastpage}


\end{document}